\documentclass[a4paper,11pt]{article}
\pdfoutput=1 

\usepackage{jinstpub} 

\usepackage{lineno}

\usepackage{booktabs, multirow} 
\usepackage{soul}
\usepackage[table]{xcolor} 
\usepackage{changepage,threeparttable} 

\title{\boldmath 100 Gb/s High Throughput Serial Protocol (HTSP) for Data Acquisition Systems with Interleaved Streaming}

\author{L. Ruckman}
\author{and D. Doering}
\affiliation{
    SLAC National Accelerator Laboratory,
    2575 Sand Hill Road, M/S 96,
    Menlo Park, CA 94025, USA
}

\emailAdd{ruckman@slac.stanford.edu}

\abstract{
Demands on Field-Programmable Gate Array (FPGA) data transport
have been increasing over the years as frame sizes and 
refresh rates increase. As the bandwidths requirements increase
the ability to implement data transport protocol layers 
using "soft" programmable logic becomes harder and start to require harden 
IP blocks implementation. 
To reduce the number of physical links and interconnects, it is common for 
data acquisition systems to require interleaving of streams on the same link
(e.g. streaming data and streaming register access). 
This paper presents a way to leverage existing
FPGA harden IP blocks to achieve a robust, low latency 100 Gb/s 
point-to-point link with minimal programmable logic overhead
geared towards the needs of data acquisition systems with interleaved streaming requirements.
}


\begin{document}
\maketitle
\flushbottom


\section{Background}
\label{sec:Background}

LCLS-II, a Free Electron Laser (FEL) X-ray light source, started operations at SLAC National Laboratory in 2020. This new machine will produce X-ray pulses with a repetition rate up to a million times per second when it completes its full commissioning. In order for the experiments to take advantage of this high repetition rate, detectors and readout systems have to operate at the same frequency and cope with the generated data volumes.

\begin{figure}[htp!]
  \centering
  \centerline{\includegraphics[width=12cm]{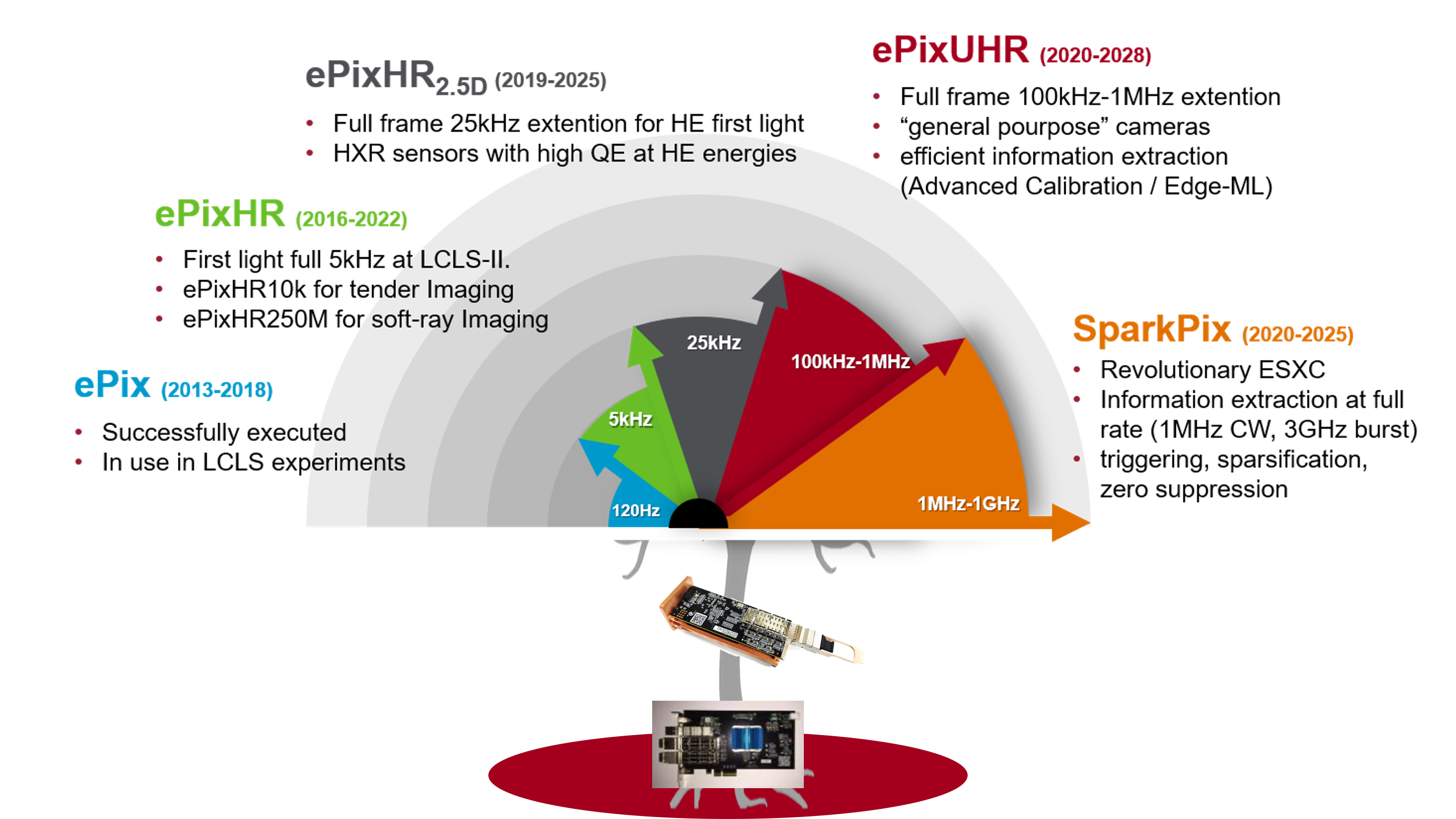}}
  \caption{ePix high rate detector family. }
  \label{fig:ePix}
\end{figure}

SLAC detector program started with the developments of cameras running at LCLS (1\textsuperscript{st} generation) with nominal rate of 120Hz. The detector development efforts were centered around the ePix ASIC family \cite{ePix2016Nishimura, van2020epix10k}. Presently a new family of high rate detectors have been proposed (see Fig. \ref{fig:ePix}), where the ePix High Rate family class of detectors are capable of frame rates of 5Kfps \cite{doeringePixHr} which is still orders of magnitude lower from the target frame rates. 

Future detector generations are already under study and Table \ref{tab:epix_datarate} summarizes the detector classes, their target rate and technology. Several design aspects are challenging in these detector development and this paper focuses on the data transmission and the impacts of such data rates have in the communication protocol used between the detector and the Data AcQuisition system (DAQs). The forecast rates of future detectors are approaching 4,000 Gbps total  data bandwidth and several parallel effort in terms of data reduction are being pursued, such as ASIC, detector and edge computing classification and compression to alleviate the data transmission challenge. Despite them it will still require multiple 100+Gpbs lanes to implement the communication of these detectors. 

Protocol for serial streaming of data at 100+Gbps exist, but they are limited in terms of implementation complexity, large resource usage (due to unnecessary features for this application or lack of use of harden IP). Due to these limitation rises the need to a new protocol that can obtain the desired speed with small resource usage by leveraging of harden IP. In this paper we introduce the 100 Gb/s High Throughput Serial Protocol (HTSP) which address the aforementioned limitations.

In the remainder of this section we provide a summary of existing protocols for data rates of 100Gbps or more. In section \ref{sec:Architecture} we describe the HTSP architecture. In section \ref{sec:Performance} the performance of the protocol is presented including bandwidth, frame rate, latency and resources. We conclude this paper in section \ref{sec:Summary} with a summary of this work.

\setlength{\arrayrulewidth}{0.1mm}
\setlength{\tabcolsep}{10pt}
\renewcommand{\arraystretch}{1.5}
\begin{table}
    \tiny
    \centering
    \caption {Expected data rate for the ePix detector family} 
    \begin{tabular}{ |p{2cm}|p{2cm}|p{2.5cm}|p{3cm}|  }
    \hline &
    \multicolumn{3}{|c|}{ePix detector parameter} \\
    \hline
    Detector class & Frame Size (MPix) & Detector frame rate (fps) & Detector data bandwidth (Gbps) \\
    \hline
    ePixHR 10K single & 0.33 & 5,000 &  10.2 \\
    ePixHR 10K TXI & 2.2 & 5,000 &  202.8 \\
    ePixHR 2.5D TXI & 2.2 & 25,000 &  1,014 \\
    ePixHR UHR & 2.2 & 100,000 &  4,056 \\
    SparkPix & 0.33 & 1,000,000 &  204*\\
    \hline
    \multicolumn{4}{l}{* considering 10\% occupancy}
    \end{tabular}
    \label{tab:epix_datarate}
\end{table}

\subsection{Interleaved Streaming}

To reduce the number of physical links and interconnects, it is common for 
DAQs to require interleaving of streams on the same link. 
An example of this would be combining the FPGA's data stream, register access stream, and other asynchronous update message streams into the same physical link.  
This requirement of stream interleaving is likely derived from the need to reduce the cabling interconnects for cost, complexity, and weight or reducing the number of vacuum penetrations.  

\subsection{Existing protocols}

There are many protocols available in the literature and commercially under free and paid licences. 
Aurora protocol\cite{aurora2014}, particularly the 64/66b, aims at streaming application on chip-to-chip transmission and can be used with latest FPGA families to obtain over 400Gbps\cite{aurora2020}. Aurora supports flow control of framed streaming data. It is a very lightweight implementation with respect to FPGA resource.  
However, it does not support interleaved streams (A.K.A. "virtual channels").  Virtual channels could be added by a custom layer on top of the Aurora IP.  But this custom layer protocol would need to support pause per virtual channel such that the Aurora flow control for 1 virtual channel does not back pressure all the other virtual channels.  This kind of virtual channel layer on top of Aurora likely would be resource intensive for LUTs, FFs and BRAMs on the FPGA. 

The Aurora IP core does support a CRC for bit error checking but does not support Forward Error Correction (FEC) in its implementation.  Typically FEC is used for high serial rates (e.g. 25.78125 Gb/s or faster) to deal with the large Bit Error Rate (BER). A DAQ application that has a requirements for high data reliability will likely need to add the FEC module.  Adding this FEC module would require patching the Aurora protocol source code due to the lack of FEC support in its implementation. 

Aurora does not support lane marker framing and de-framing including reordering of each lane, which is another draw back. Aurora requires that lane(0) maps to lane(0) on both side, lane(1) maps to lane(1) on both sides, etc. Lane markers for framing, de-framing and lane reordering is a feature supported in 100 Ethernet PHY.  This means 100 Ethernet is robust against common lane reversals that often happen when connecting through optical patch panels between the two end points.

Aurora 64/66b protocol is the closest existing protocol that meets our X-ray cameras requirements (point-to-point communication and low FPGA resource utilization). Aurora protocol does not support interleaving of streams, which was one of the motivations for developing HTSP.
\section{Architecture}
\label{sec:Architecture}

\begin{figure}[tb]
\centering
\includegraphics[width=1.0\textwidth]{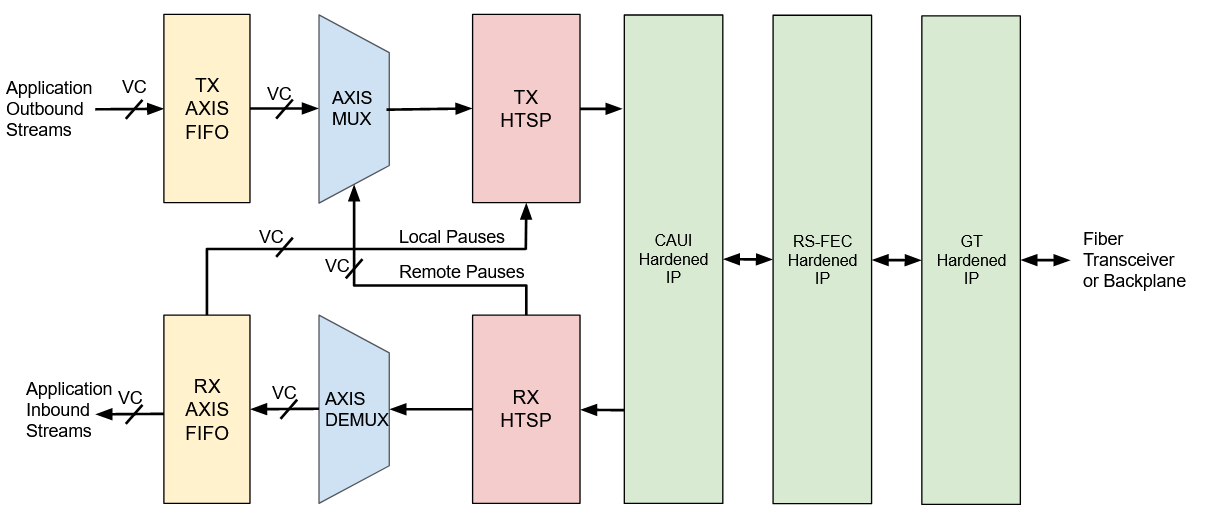}
\caption{\label{fig:architecture_block_diagram}
HTPS Firmware Architecture Block Diagram}
\end{figure}

A block diagram of the HTSP firmware core 
is shown in Figure \ref{fig:architecture_block_diagram}.  
The HTSP firmware is composed of three parts: 
Harden IP cores, HTSP protocol layer, and application interface.
All streaming interfaces inbound and outbound of HTSP use
Advanced eXtensible Interface Stream (AXIS) protocol\cite{AXIS_SPEC}.
For this 100 Gb/s implementation, these AXIS buses are all 512-bit wide operating at 195.66 MHz.
HTSP supports up to 16 Virtual Channels (VCs) streams.
Each VC stream at the application layer can be treated as an independent AXIS stream.
When the VC streams are combined together with an AXIS multiplexer (MUX),
the AXIS's TDEST metadata field is used to encode the VC stream index.
When the VC stream from HTSP RX is slip apart with an AXIS demultiplexer (DEMUX),
the AXIS's TDEST metadata field is used to decode the VC stream index.
Each VC has a RX AXIS FIFO with a "pause" threshold set to a 
fraction of total FIFO buffering depth.
This FIFO is configured to be 
large enough to absorb any data in-flight once the pause is asserted.

There are two types of pauses: Local and remote.  Local pauses are 
driven by the local VC RX FIFOs and published on the HTSP TX outbound frames.
The remote pauses are received by the RX HTSP inbound frames and used 
by the AXIS MUX to determine if a given VC needs to halt data transport.
By having independent pauses per VC stream, back pressure from one VC 
does not prevent all other VC streams from moving their payload data across the link.

HTPS supports interleaving of application AXIS streams on the HTPS link.  
The AXIS MUX can send a partial AXIS frame for a given VC stream
(up to a programmable limit) then arbitrate to the next VC stream.
The header and footer (refer to Section \ref{sec:protocol}) contain 
all the metadata information for combining these interleaving partial frames back together.

\subsection{Harden IP Cores}

To minimize the amount of programmable logic resources 
(e.g. Look Up Tables (LUTs), Flip Flops (FFs), etc), 
HTSP leverages the existing FPGA integrated harden IP modules.
There are three harden IP modules required for HTSP implementation: CAUI, RS-FEC, and GT.
CAUI is the 100 Gigabit Attachment Unit Interface module. 
Some of the key CAUI features that HTSP takes advantage of are:
\begin{itemize}
   \item Frame Check Sequence (FCS) calculation and addition in the transmit (TX) direction
   \item FCS checking and FCS removal in the receive (RX) direction
   \item Physical Coding Sublayer (PCS) lane marker framing and de-framing 
   \begin{itemize}
        \item Including reordering of each PCS lane
    \end{itemize}
\end{itemize}
The Reed-Solomon Forward Error Correction (RS-FEC) layer provides the ability to correct
for multiple bit errors on the serial link.
The Gigabit Transceivers (GT) layer handles the serialization/deserialization. The GTs can be configured for either
4-lane mode (CAUI-4, 25.78125 Gb/s/lane) or 10-lane mode (CAUI-10, 10.3125 Gb/s/lane).  
FPGA manufactures (e.g. Xilinx\cite{xilinx_caui} and Intel\cite{Intel_caui}) provide 
IP wizards that contain all threee harden IP modules and help with their complex 
configurations.


\subsection{HTPS Protocol}
\label{sec:protocol}

The HTPS frame is made of either a header only or a (header + payload + footer).
Header only is used to establish the link and publish the updated pause status 
to the other side of the link. The header only frame occurs based on a periodic 
timer.  This timer is reset when a full frame (header + payload + footer) is sent.
The payload is the user data being sent across the link.

The header format is show in Table \ref{tab:header}.
The header's first 14 bytes are the standard Ethernet 
Destination/Source MAC and EtherType.  
This make the HTSP capable of operating on standard Ethernet switches and routers
by formatting the frame as a custom RAW Ethernet frame.  However for DAQ,
the preference is likely to only operate in point-to-point to 
reduce/minimize the latency and remove the requirement to add a 
reliability layer due to drop potential buffer overflows in the Ethernet switches/routers.
"Version" is a hard coded value to 0x1, which is checked as part of establishing link.
TID is the frame's transaction ID and is a counter that increments by one per TX frame.
TID is used for debugging. 
For the HTPS TX module, "pause" is current state of the \textbf{\textit{local}} RX FIFOs.
For the HTPS RX module, "pause" is the current state of the \textbf{\textit{remote}} RX FIFOs.
VC is the index of the virtual channel stream being sent on the payload.
TUserFirst is the payload's first 8-bit of the AXIS's TUSER metadata field.
TUserFirst is use to pass metadata on the payload frame (e.g. Start of the payload frame).
OpCodeEn is the valid flag for the 128-bit operation-code Data (OpCodeData).
UserData is a 128-bit bus that is sampled every header.
UserData can be used to publish asynchronous status updates 
or board information (e.g. serial number) to the 
other side of the link.  There are 10 bytes that are reserved and set to zero.  
HdrXsum is a 16-bit checksum for only the header and not the payload or footer.

The header and payload are both match to 512-bits.  While it would be 
possible to make the header less than 512-bits because there are 10 bytes reserved,
the additional programmable logic for byte packing would increase the resource utilization
and increase the HTSP programmable logic's complexity. 
For large HTPS frames (e.g. 8kB payloads), the reduced bandwidth 
efficiency for a slightly increased header size is minimal. 

\begin{table}[!htb]\centering
\caption{HTSP Header Format}\label{tab:header}
\begin{tabular}{|c|l|l|}
\hline
Byte Index &Name &Description \\ \hline\hline
5:0     & DestMac       & Destination MAC \\ \hline
11:6    & SrcMac        & Source MAC \\ \hline
13:12   & EtherType     & Ethernet Type \\ \hline
14      & Version       & Version=0x1 \\ \hline
15      & TID           & Transaction ID \\ \hline
17:16   & Pause         & Virtual Channel Pause \\ \hline
18      & VC            & Virtual Channel Index \\ \hline
19      & TUserFirst    & first 8-bits of AXIS TUSER \\ \hline
20      & OpCodeEn      & OP-code Enable \\ \hline
29:20   & Reserved      & Zeros \\ \hline
31:30   & HdrXsum       & 16-bit Header Checksum \\ \hline
47:32   & OpCodeData    & 128-bit OP-code Data \\ \hline
63:48   & UserData      & 128-bit User Data \\ \hline
\end{tabular}
\end{table}

The footer format is shown in Table \ref{tab:footer}.
The footer is only 6 bytes (48-bit).
TKeepLast is the AXIS TKEEP (A.K.A. "byte enable") on the last payload word in the HTSP frame.
This provides support for transporting payloads less than 64 bytes (512-bit).
TLast/TUSER is mapped to the 2nd byte of the footer. 
TLast/TUSER.BIT(0) is the payload's TLAST value (A.K.A. "end of frame").
TLast/TUSER.BIT(7:1) is the payload's last upper 7-bits of AXIS TUSER with respect to the last byte.
This TUSER metadata is used to pass metadata on the payload frame (e.g. End of payload frame with Error).
The footer "pause" is the latched pause value from the header during the payload until the footer ends and it is a latched value in case the pause asserts during the payload transport. 
PayloadSize is the number of byte in the payload and checked by the HTPS RX module for errors.

There is no Cyclic Redundancy Check (CRC) field in the footer.
This is because HTPS utilizes the harden IP's FCS that is 
calculated/inserted in TX direction and checked/removed in the RX direction.
We mapped harden IP's FCS error flag into harden IP's outbound AXIS TUSER field
and is checked by the HTPS RX module.

\begin{table}[!htb]\centering
\caption{HTSP Footer Format}\label{tab:footer}
\begin{tabular}{|c|l|l|}
\hline
Byte Index &Name &Description \\ \hline\hline
0       & TKeepLast     & AXIS TKEEP on last payload word \\ \hline
1       & TLast/TUSER   & AXIS TLAST and last 7-bits of AXIS TUSER  \\ \hline
3:2     & Pause         & Virtual Channel Pause \\ \hline
5:4     & PayloadSize   & Number of bytes in payload \\ \hline
\end{tabular}
\end{table}

\section{Performance}
\label{sec:Performance}

\begin{figure}[tb]
\centering
\includegraphics[width=0.3\textwidth]{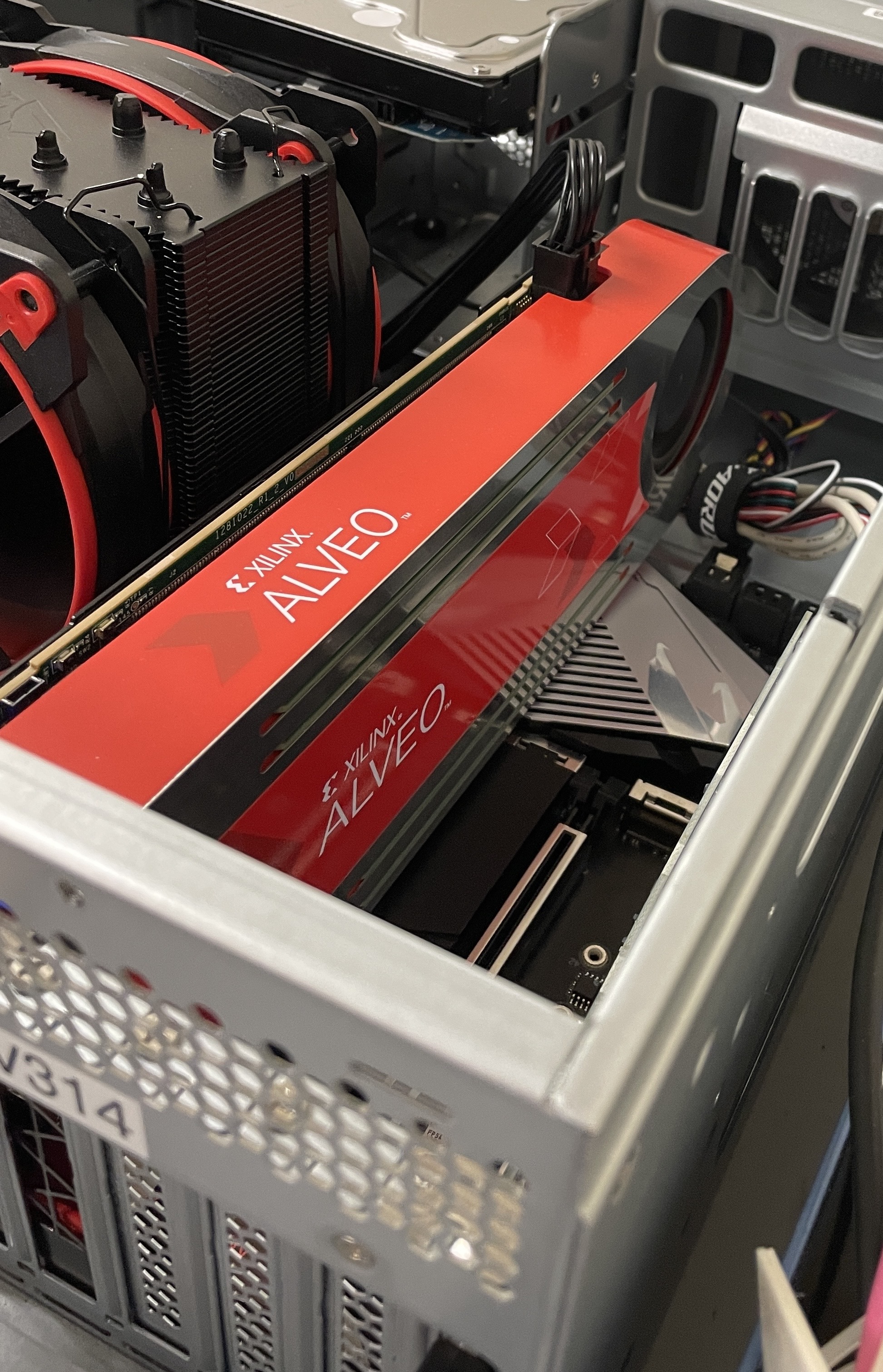}
\caption{\label{fig:au200_image} Photograph of Xilinx Alveo U200 Data Center Accelerator Card installed into a server}
\end{figure}

In this section we present the performance test results. The Xilinx U200 PCIe card\cite{Alveo_U200} was used for the testbench physical implementation.
A photograph of this PCIe card is shown in Figure \ref{fig:au200_image}. 
This PCIe card was selected for these measurements because its FPGA supports both 25Gb/s/lane on the QSFP28 interface (CAUI-4 mode) and included the integrated "100G Ethernet with RS-FEC" harden IP.

\begin{figure}[tb]
\centering
\includegraphics[width=0.7\textwidth]{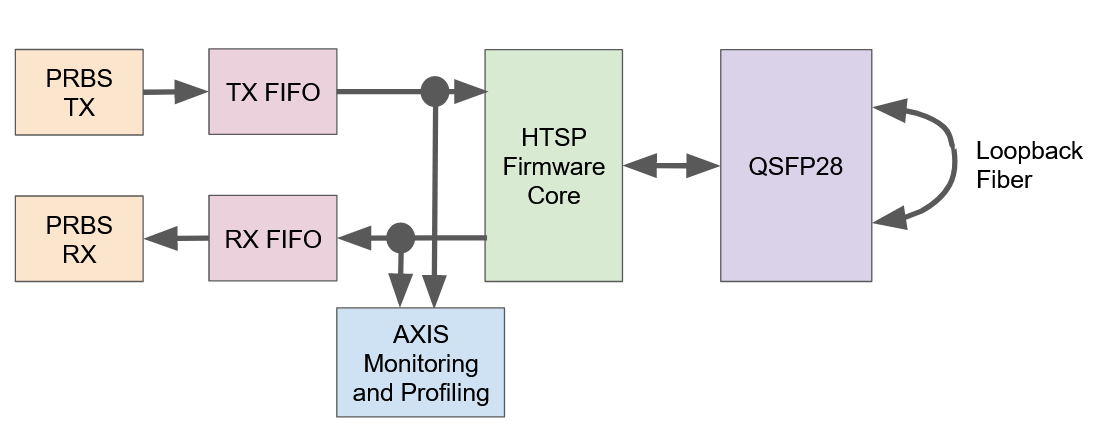}
\caption{\label{fig:perf_fw_block_diagram} 
Block diagram of the performance/benchmark}
\end{figure}

A block diagram of the performance/benchmark firmware is shown in Figure \ref{fig:perf_fw_block_diagram}.  
A Pseudo Random Binary Sequence (PRBS) Transmitter (TX) generates AXIS frames that feeds into the HTSP firmware. The PRBS TX is able to generate programmable AXIS frame sizes and is able to generate frames faster than the 100 Gb/s serial link.
The HTSP firmware module is configured for 8kB payload burst size.
The HTSP's output is loopbacked to itself using a fiber optic cable on the QSFP28 pluggable module.
The HTSP firmware forwards the received stream 
to the PRBS receiver (RX).  The PRBS RX firmware module checks for any bit errors in the AXIS frame.  

An AXIS profiling module monitors the HTSP's inbound/outbound AXIS streams and provides information such has average bandwidth and average frame rate over a 1 second refresh rate. This profiling module also includes the ability to measure the latency between the start of an inbound frame to the start of an outbound frame for single one-shot single payload frame measurements. 

\subsection{Bandwidth}
\label{sec:Bandwidth}
\begin{figure}[tb]
\centering
\includegraphics[width=1.0\textwidth]{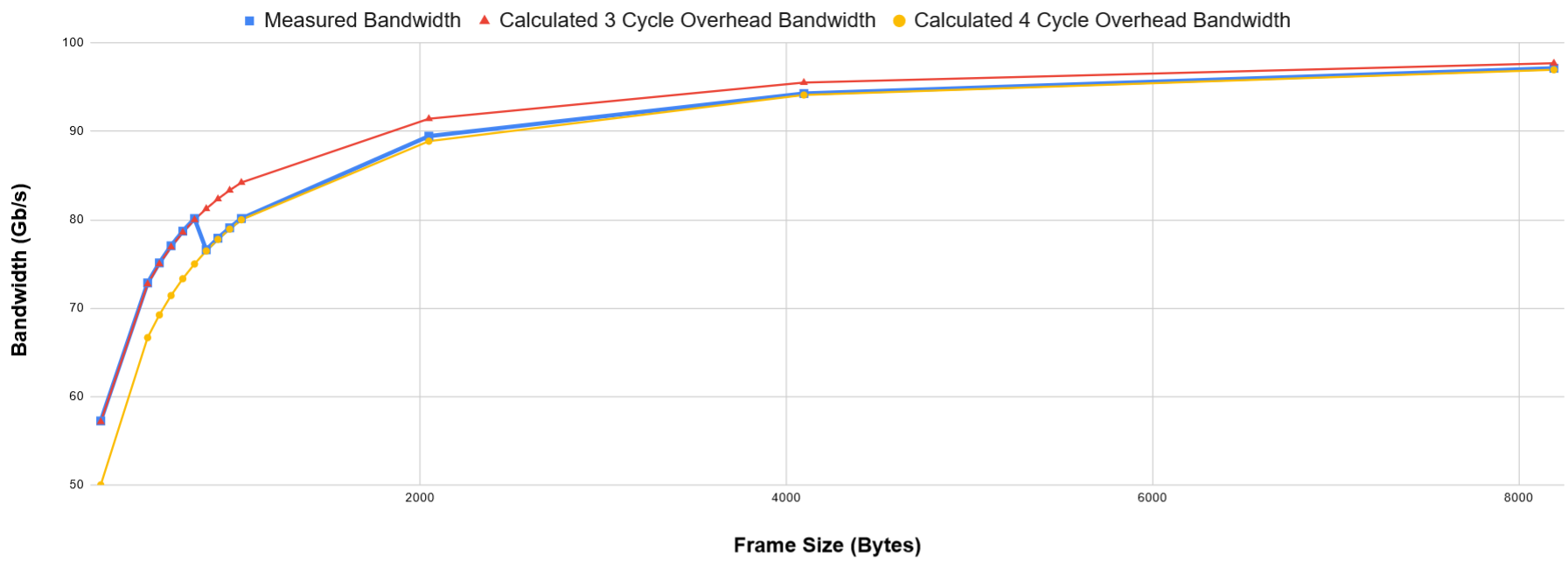}
\caption{\label{fig:bandwidth_vs_framesize}
Plot of HTSP's bandwidth versus frame size}
\end{figure}

The results of the bandwidth measurements are shown in Figure \ref{fig:bandwidth_vs_framesize}.  
The AXIS profiling firmware module provided the measured bandwidth values.
The calculated bandwidth is the following:

\begin{align*}
 bandwidth = \frac{ (100 Gb/s)*(frame\ size)}{(frame\ size) + (overhead)}
\end{align*}

The expected overhead was a constant 3 clock cycle (192 bytes) for the HTSP header, HTSP footer, and Ethernet Interpacket Gap (IPG). We observed that the calculated 3 cycle overhead bandwidth and measured bandwidth agreed fairly well up to 768 bytes frame size.  When the AXI stream frame size became larger than 768 bytes, the measured bandwidth was tracking a calculated bandwidth with 4 cycle overhead.  Likely the harden IP is asserting an additional overhead cycle for larger frame sizes. 

Aurora 64/66b protocol overhead is 1 clock cycle for every 32 clock cycles of data (64 data bytes per cycle). This is a 3 percent inefficiency for transmission overhead (or 97 percent bandwidth efficiency).  For less than 8kB frame sizes, Aurora 64/66b protocol has better bandwidth efficiency than HTSP.  But for 8KB (or larger) frames, HTSP is the same 97 bandwidth efficiency as Aurora 64/66b protocol.

\subsection{Frame Rate}

\begin{figure}[tb]
\centering
\includegraphics[width=1.0\textwidth]{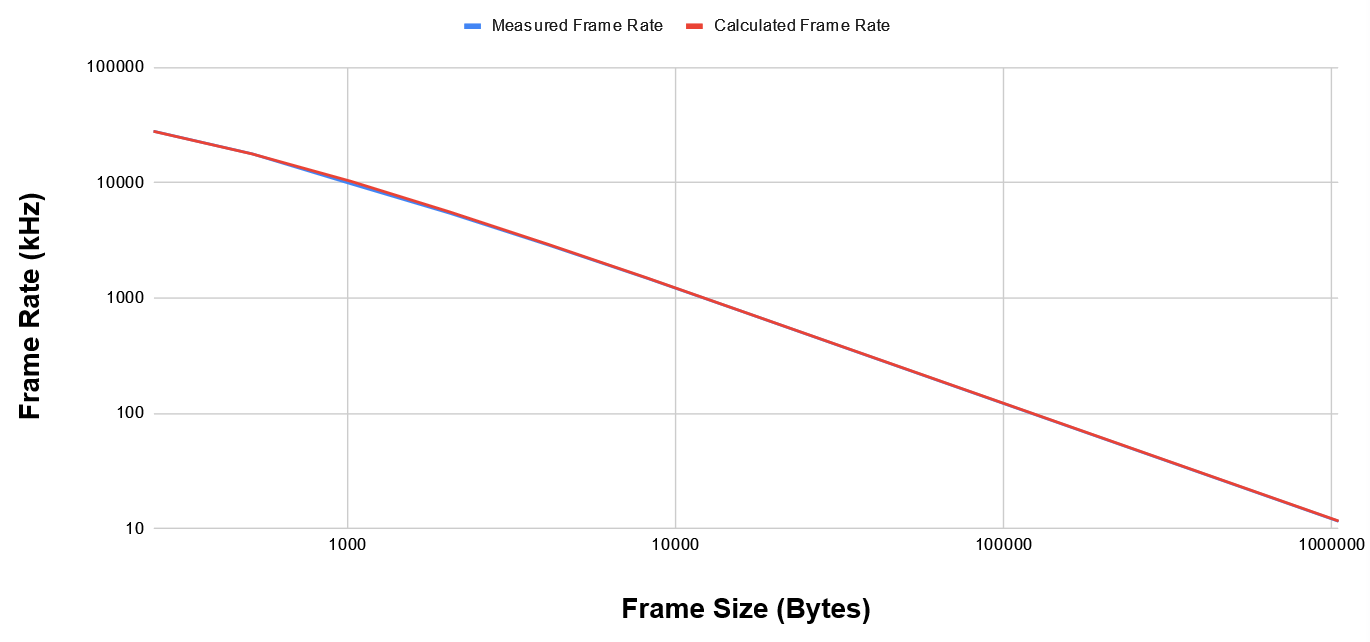}
\caption{\label{fig:framerate_vs_framesize}
Plot of HTSP's frame rate versus frame size}
\end{figure}

The results of the frame measurements are shown in Figure \ref{fig:framerate_vs_framesize}.  
The AXIS profiling firmware module provided the measured frame rate values.
The calculated frame rate is the following:

\begin{align*}
 frame\ rate = \frac{ (100 Gb/s)}{(frame\ size) + (overhead)}
\end{align*}

We observed that the calculated frame rate and frame rate agree fairly well.  
For small 256B frames, the measured frame rate is 28.0 MHz 
and calculated frame rate  is 27.9 MHz.  
For large 1MB frames, the measured frame rate is 11.6 kHz 
and calculated frame rate  is 11.6 kHz.  

\subsection{Latency}

\begin{figure}[tb]
\centering
\includegraphics[width=1.0\textwidth]{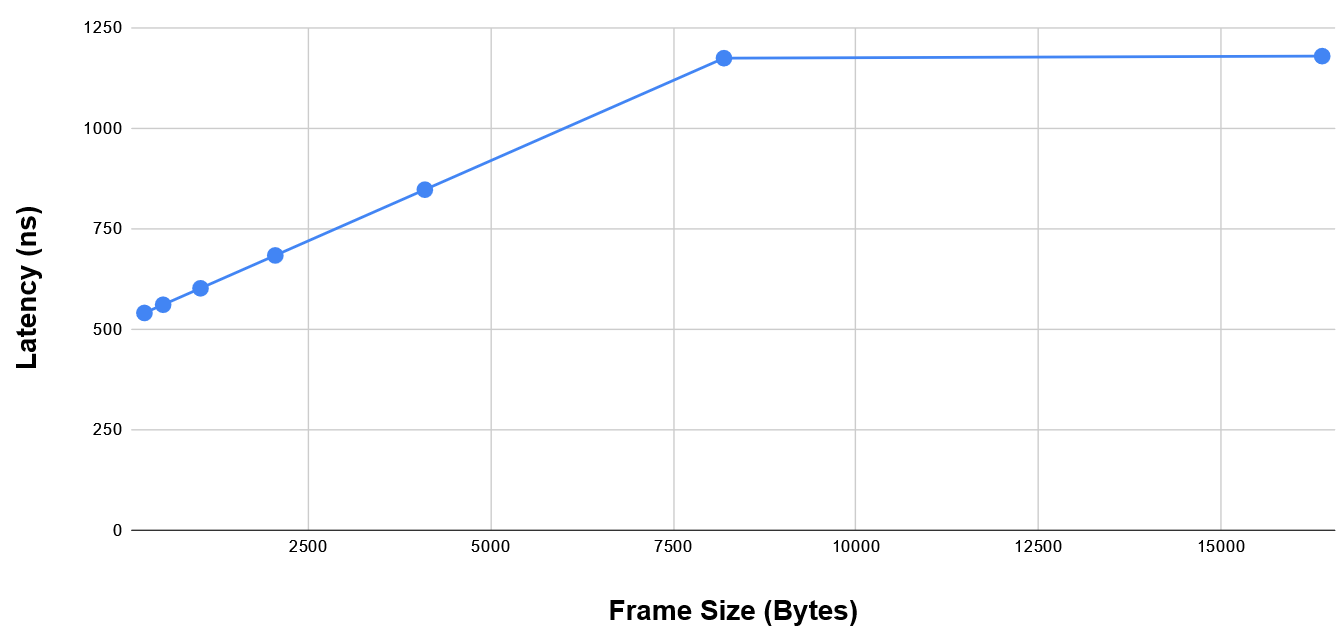}
\caption{\label{fig:framerate_vs_latency}
Plot of HTSP's latency versus frame size}
\end{figure}

The results of the latency measurements are shown in Figure \ref{fig:framerate_vs_latency}.  
The latency is the time between the start of an inbound AXIS frame into the HTPS TX module
to the start of an outbound AXIS frame out of HTSP RX module.
The AXIS profiling firmware module provided this measured latency values.
The linear ramp up to 8kB is due to a "store-and-forward" FIFO in between
the HTSP and CAUI4 harden IP core to prevent AXIS TVALID from de-asserting 
while in the middle of a frame transport.
Because the firmware is configured for 8kB payload burst size, 
the latency is constant after 8kB frames, which is 1.176 microsec.

\subsection{FPGA Resources}

HTSP resources depend on the number of Virtual Channels that are being implemented at a given setup. In order to make a fair comparison with AURORA 6466b Table \ref{tab:FPGA_Resources_Aurora} presents the HTSP firmware resources for only 1 Virtual Channel versus the equivalent Aurora protocol.  The RAM utilization is higher in HTSP because the FIFOs for streaming are included with the protocol, where Aurora protocol does not. 
HTSP protocol is leveraging the DSP slices in the interleaving, where Aurora protocol does not. 
For a single Virtual Channel configuration, as expected, due to the reduced functionality Aurora 64b66b is significantly lower in resources.

\begin{table}[!htp]\centering
\caption{Firmware resource utilization for HTSP With only 1 VC versus Aurora 64/66b protocol at 4 lanes x 25Gbps}\label{tab:FPGA_Resources_Aurora}.
\tiny
\begin{tabular}{|c|c|c|c|c|c|c|}
\hline
\textbf{\# of VCs} & \textbf{LUT(K)} & \textbf{FF(K)} & \textbf{BRAM36} & \textbf{URAM} & \textbf{DSP48} \\ \hline
\hline
HTSP(VC=1)    & 19.3 & 43.4 & 32.5 & 8  & 16\\ \hline
Aurora 64b66b & 11.2 &  9.5 &    6 & 0  & 0\\ \hline
\end{tabular}
\end{table}

A key feature of HTSP is its capability of interleaving streams based on the virtual channels concept. Table \ref{tab:FPGA_Resources} presents the HTSP firmware resources as a function of Virtual Channels (VCs).
To optimize the firmware logic, the number of VCs is a build time VHDL generic.
When using the minimal interleaving of streaming data option, 
then building the HTSP firmware with only 2 VC configuration uses just over 2\% of the U200 resource utilization and it increased to only 8\% for 16 VCs.
The very low FPGA resource utilization is possible by leveraging the harden IP features as much as possible.

\begin{table}[!htp]\centering
\caption{HTSP FPGA firmware resource utilization, in parentheses is U200 utilization}\label{tab:FPGA_Resources}.
\tiny
\begin{tabular}{|c|c|c|c|c|c|c|}
\hline
\textbf{\# of VCs} & \textbf{LUT(K)} & \textbf{FF(K)} & \textbf{BRAM36} & \textbf{URAM} & \textbf{DSP48} \\ \hline
\hline
1 &19.3 (1.6\%) &43.4 (1.8\%) &32.5 (1.5\%) &8 (0.8\%) &16 (0.2\%) \\ \hline
2 &24.9 (2.1\%) &55 (2.3\%) &32.5 (1.5\%) &16 (1.7\%) &28 (0.4\%) \\ \hline
3 &30.2 (2.6\%) &66.4 (2.8\%) &32.5 (1.5\%) &24 (2.5\%) &40 (0.6\%) \\ \hline
4 &36.2 (3.1\%) &78 (3.3\%) &32.5 (1.5\%) &32 (3.3\%) &52 (0.8\%) \\ \hline
5 &40.4 (3.4\%) &89.5 (3.8\%) &32.5 (1.5\%) &40 (4.2\%) &64 (0.9\%) \\ \hline
6 &46 (3.9\%) &101.1 (4.3\%) &32.5 (1.5\%) &48 (5\%) &76 (1.1\%) \\ \hline
7 &50.6 (4.3\%) &112.6 (4.8\%) &32.5 (1.5\%) &56 (5.8\%) &88 (1.3\%) \\ \hline
8 &55.1 (4.7\%) &124 (5.2\%) &32.5 (1.5\%) &64 (6.7\%) &100 (1.5\%) \\ \hline
9 &61.4 (5.2\%) &135.7 (5.7\%) &32.5 (1.5\%) &72 (7.5\%) &112 (1.6\%) \\ \hline
10 &66.4 (5.6\%) &147.2 (6.2\%) &32.5 (1.5\%) &80 (8.3\%) &124 (1.8\%) \\ \hline
11 &72.6 (6.1\%) &158.8 (6.7\%) &32.5 (1.5\%) &88 (9.2\%) &136 (2\%) \\ \hline
12 &77.2 (6.5\%) &170.3 (7.2\%) &32.5 (1.5\%) &96 (10\%) &148 (2.2\%) \\ \hline
13 &82 (6.9\%) &181.9 (7.7\%) &32.5 (1.5\%) &104 (10.8\%) &160 (2.3\%) \\ \hline
14 &87.4 (7.4\%) &193.4 (8.2\%) &32.5 (1.5\%) &112 (11.7\%) &172 (2.5\%) \\ \hline
15 &92.9 (7.9\%) &204.9 (8.7\%) &32.5 (1.5\%) &120 (12.5\%) &184 (2.7\%) \\ \hline
16 &95.7 (8.1\%) &216.6 (9.2\%) &32.5 (1.5\%) &128 (13.3\%) &196 (2.9\%) \\ \hline
\end{tabular}
\end{table}

\section{Summary}
\label{sec:Summary}

Novel scientific experiments where high data transmission rates are needed impose challenges to the first level FPGA. This is particularly the case for the next generation x-ray cameras required by LCLS-II where frame rates approach 1MHz and total data thought put approach 4Tbps. The state of the art electronics system requires multiple parallel links to achieve such performances. The performance of each of this links must be optimized, including the protocol definition and its resource usage.

In this paper we presented HTSP as an efficient alternative to existing protocols that leverages from harden IP, therefore minimizing the use of programmable logic for interleaved communication. Measurements were made of the bandwidth, latency and programmable logic utilization. For example, the measured bandwidth obtained with HTSP was 97.1Gb/s (4 lanes) where the calculated bandwidth was 97.7, achieving a 99.4\% efficiency. HTSP firmware has been developed and released on Github\cite{slaclab_surf}.

For a single virtual channel configuration,  Aurora 64/66b protocol is clearly the better option with respect to bandwidth efficiency and FPGA firmware resource.  But for a multiple virtual channel configuration, HTPS requires less custom user logic than Aurora 64/66b protocol because the stream channelization is embedded in the HTPS protocol. 

The performance obtained by the HTSP when implemented in FPGA has demonstrated that it meets the requirements for next generation of ePix high rate detectors.

\end{document}